\documentclass[twocolumn,prl,aps]{revtex4}
\usepackage{graphicx}
\usepackage{bm}
\usepackage{color}

\def\etal{\textit{et al.}}

\begin{document}
\title{Reply to: Comment by S.-L. Drechsler \etal\ (arXiv:1006.5070v1)}

\author{M. Enderle}
\affiliation{Institut Laue Langevin, BP156, 6 rue Horowitz, 38042
Grenoble, France}
\author{B. F{\aa}k}
\affiliation{Commissariat \`a l'Energie Atomique, INAC, SPSMS, 38054
Grenoble, France}
\author{H.-J. Mikeska}
\affiliation{Institut f\"ur Theoretische Physik, Leibniz
Universit\"at Hannover, Appelstrasse 2, 30167 Hannover, Germany}
\author{R. K. Kremer}
\affiliation{Max-Planck Institute for Solid State Research,
Heisenbergstrasse 1, 70569 Stuttgart, Germany}

\begin{abstract}
In a comment on arXiv:1006.5070v1, Drechsler \etal\ present new
band-structure calculations suggesting that the frustrated
ferromagnetic spin-1/2 chain LiCuVO$_4$ should be described by a
strong rather than weak ferromagnetic nearest-neighbor interaction,
in contradiction with their previous calculations. In our reply, we
show that their new results are at odds with the observed magnetic
structure, that their analysis of the static susceptibility neglects
important contributions, and that their criticism of the spin-wave
analysis of the bound-state dispersion is unfounded. We further show
that their new exact diagonalization results reinforce our
conclusion on the existence of a four-spinon continuum in
LiCuVO$_4$, see Enderle \etal, Phys. Rev. Lett. 104 (2010) 237207.
\end{abstract}

\maketitle

In their Comment, Drechsler \etal\ \cite{Drechsler} argue that the
frustrated ferromagnetic spin-1/2 chain LiCuVO$_4$ should be
described by a {\it strong} ferromagnetic (FM) nearest neighbor (NN)
interaction $J_1$, rather than a {\it weak} FM $J_1$ as found by an
analysis of the static susceptibility and the dispersion of the
bound state \cite{Enderle05} and also from an analysis of the
intensity of the continuum scattering \cite{Enderle10}. Drechsler
\etal\ have in \cite{Drechsler} revised their band-structure
calculations of \cite{Enderle05} and now propose two alternative
sets of exchange integrals with strongly FM $J_1$, which are claimed
to be in [better] agreement with experiment. The set A has
$J_1=-6.3$ and $J_2=5.05$ meV while set B has $J_1=-8.8$, $J_2=6.5$,
and an antiferromagnetic (AFM) interchain coupling (cf.
\cite{Enderle05}) $J_4=0.5$ meV.

We first note that the large AFM $J_4$ of set B leads to a
propagation vector with a component along the $a^\star$ axis, in
contradiction with the observed magnetic structure.

\begin{figure}[b]
\includegraphics[width=.97\columnwidth]{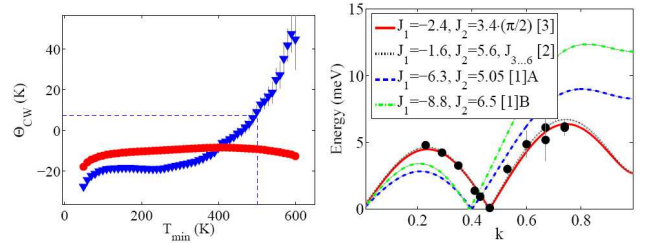}
\caption{(a) Curie-Weiss temperature $\theta_{CW}$ as a function of
the lowest temperature $T_{\rm min}$ used for fits of the inverse
susceptibility. (The highest temperature is always 640 K). (Blue)
triangles: Procedure of \cite{Drechsler}. (Red) circles: Fits after
correction for van-Vleck and diamagnetic contributions with $g_a$
fixed to ESR value \cite{Nidda02}. (b)~Chain dispersion compared to
spin-wave calculations with different exchange parameters.}
\end{figure}

Secondly, the inverse static susceptibility of \cite{Enderle05} was
reanalyzed in \cite{Drechsler} by fitting a linear relation to the
high-temperature part above 500 K, with the $g$-factor and
$\theta_{CW}$ as free parameters. Figure 1a shows the Curie-Weiss
temperature $\theta_{CW}$ from such linear fits of the same data
\cite{Enderle05} as a function of the lowest temperature $T_{\rm
min}$ used (triangles). Clearly, the cut-off at $T_{\rm min}$=500 K
used by Drechsler \etal\ \cite{Drechsler} (dashed line) is fully
arbitrary and their $\theta_{CW}$ is not unique. If the
susceptibility is corrected for the known constant diamagnetic and
Van-Vleck susceptibilities and the $g$-factor fixed to the value
precisely known from ESR \cite{Nidda02}, $\theta_{CW}$ is rather
independent of the fit interval (circles) and always negative, in
agreement with the exchange integrals of \cite{Enderle05,Enderle10},
but not with set A or B of \cite{Drechsler}. Additionally, our
exchange integrals, where $J_1$ is weakly FM, agree with the more
sophisticated analysis of the static susceptibility performed in
\cite{Enderle05}, which included high-temperature series expansion
and $N$=16-ring calculations.

Thirdly, Drechsler \etal\ further argue that it is not possible to
find a unique set of $J_1$ and $J_2$ from the dispersion of the main
peaks in inelastic neutron scattering (INS) data. Within spin-wave
(SW) theory, this statement is incorrect. Figure 1b shows that the
SW dispersion along the chain direction depends strongly on the
choice of $J_1$ and $J_2$. We note that the red curve shown in Fig.
1b,c of \cite{Drechsler} does not correspond to the SW description
\cite{Enderle05}.

The exact diagonalization \cite{Drechsler} reveals that the
intensity maximum at energy $E_m/J_2$ and $k \approx 3\pi$/4 varies
little between $J_1/J_2=-0.45$ and $-1.25$ ($E_m/J_2\!\approx\!1.67$
and 1.41). If we use this exact diagonalization result to estimate
$J_2$ directly from the experimental value $E_m\!=\!6.2$ meV we find
$J_2$=3.8--4.4 meV in agreement with \cite{Enderle05,Enderle10}.
$J_2$ as large as 5 meV or even 6.5 meV can clearly be excluded. The
INS data reveal considerable intensity above an upper two-spinon
boundary $\pi J_2 \sin k$ calculated with even the largest $J_2=4.4$
meV. Interestingly, such intensity above $\pi J_2 \sin k$ is also
visible in the exact diagonalization results for both weak
($J_1/J_2=-0.45$) and strong ($J_1/J_2=-1.25$) FM coupling
\cite{Drechsler}. This intensity is in both cases much stronger than
in the AFM Heisenberg chain \cite{Caux06}. For weak FM coupling,
this confirms four-spinon excitations above a two-spinon continuum,
and reinforces our conclusion \cite{Enderle10} on the existence of a
four-spinon continuum in LiCuVO$_4$.

In our RPA model of two coupled Heisenberg chains \cite{Enderle10} a
significant modification of the isolated Heisenberg chain two-spinon
continuum is found for FM coupling between the chains, in good
agreement with the main experimental features. The fit of both $J_1$
and $J_2$ leads to a quantitatively satisfactory overall description
of the observed intensity with $|J_1|\!<\!|J_2|$ and values close to
those found in \cite{Enderle05}. The RPA-approach appears therefore
justified, at least {\sl a posteriori}.

In conclusion, the parameter sets suggested in the Comment
\cite{Drechsler} are at odds with several experimental properties,
such as the susceptibility, the magnetic ordering vector for one of
the parameter sets, and the observed dispersion of the intensity
maxima \cite{Enderle05}, while the parameters proposed in
\cite{Enderle05,Enderle10} lead to a consistent description of these
properties.

\end{document}